\documentclass[3p,epsfig,times]{elsarticle}

\usepackage{ecrc}
\usepackage{graphicx}

\volume{00}

\firstpage{1}

\journalname{Nuclear Physics A}

\runauth{R. Vogt {\it et al.}}

\jid{procs}

\jnltitlelogo{Nuclear Physics A}

\usepackage{amssymb}

\usepackage[figuresright]{rotating}

\begin{document}

\begin{frontmatter}



\dochead{}

\title{Improving the $J/\psi$ Production Baseline at RHIC and the
LHC}


\author{R. Vogt$^{1,2}$, R. E. Nelson$^{1,2}$ and A. D. Frawley$^{3}$}

\address{$^1$Lawrence Livermore National Laboratory, Livermore, CA 94550, 
USA 
\\ $^2$University of California, Davis, CA 95616, USA 
\\ $^3$Florida State University, Tallahassee, FL 32301, USA}

\begin{abstract}
We assess the theoretical uncertainties on the inclusive $J/\psi$ production
cross section in the Color Evaporation Model (CEM) using values for the 
charm quark mass, renormalization and factorization scales obtained
from a fit to the charm production data.  We use our new results to provide
improved baseline comparison calculations at RHIC and the LHC.  We also study 
cold matter effects on $J/\psi$
production at leading relative to next-to-leading order in the CEM within this
approach.
\end{abstract}


\end{frontmatter}



Because the charm quark mass is finite, the total charm production cross
section can be calculated in perturbative QCD.  However, there are large 
uncertainties due
to the choice of quark mass, factorization scale and renormalization scale
\cite{RVjoszo}.
Typical lower limits of the factorization and
renormalization scales are half the chosen charm quark mass \cite{RVjoszo,CNV}.
In this case, the
factorization scale is below the minimum scale of the parton densities.
In addition, for renormalization scales below 1 GeV, the strong coupling
constant $\alpha_s$ becomes large and the perturbative expansion is unlikely 
to converge.  Thus we seek a set of physically defensible
mass and scale parameters that reduce the cross section uncertainty.
Because the $J/\psi$ cross
sections are calculated with the same set of mass and scale parameters as
open charm production in the Color Evaporation Model \cite{Gavai:1994in}, we
also place limits on the $J/\psi$ cross section calculated in the 
CEM for the first time.  See Ref.~\cite{NVF} for full details.

The charm quark mass we employ in our calculations is the Particle Data Group 
(PDG) value based on lattice
determinations of the charm quark mass in the $\overline{\rm MS}$ scheme
at $\mu = m$: $m(m) = 1.27 \pm 0.09$~GeV
\cite{latticemass}.  We fit the factorization and renormalization 
scale parameters to a subset of the fixed target total charm production
data with $250 \leq E_{\rm beam} \leq 920$ GeV.
The data were evaluated and adjusted to the values we employ
in our fits
using the most up-to-date branching ratios for the measured decay channels
in Ref.~\cite{CarlosHermine}.
We also include data from both PHENIX \cite{PHENIX200} and STAR 
\cite{STAR11,STAR11final} at $\sqrt{s_{_{NN}}} = 200$ GeV.
We neglect unknown
next-order uncertainties which could be large for charm where the mass
is relatively small and ${\mathcal O}(\alpha_s^4)$ corrections could be 
significant.  

\begin{figure}[htb]
\begin{center}
\includegraphics[width=0.245\textwidth,height=0.25\textheight]{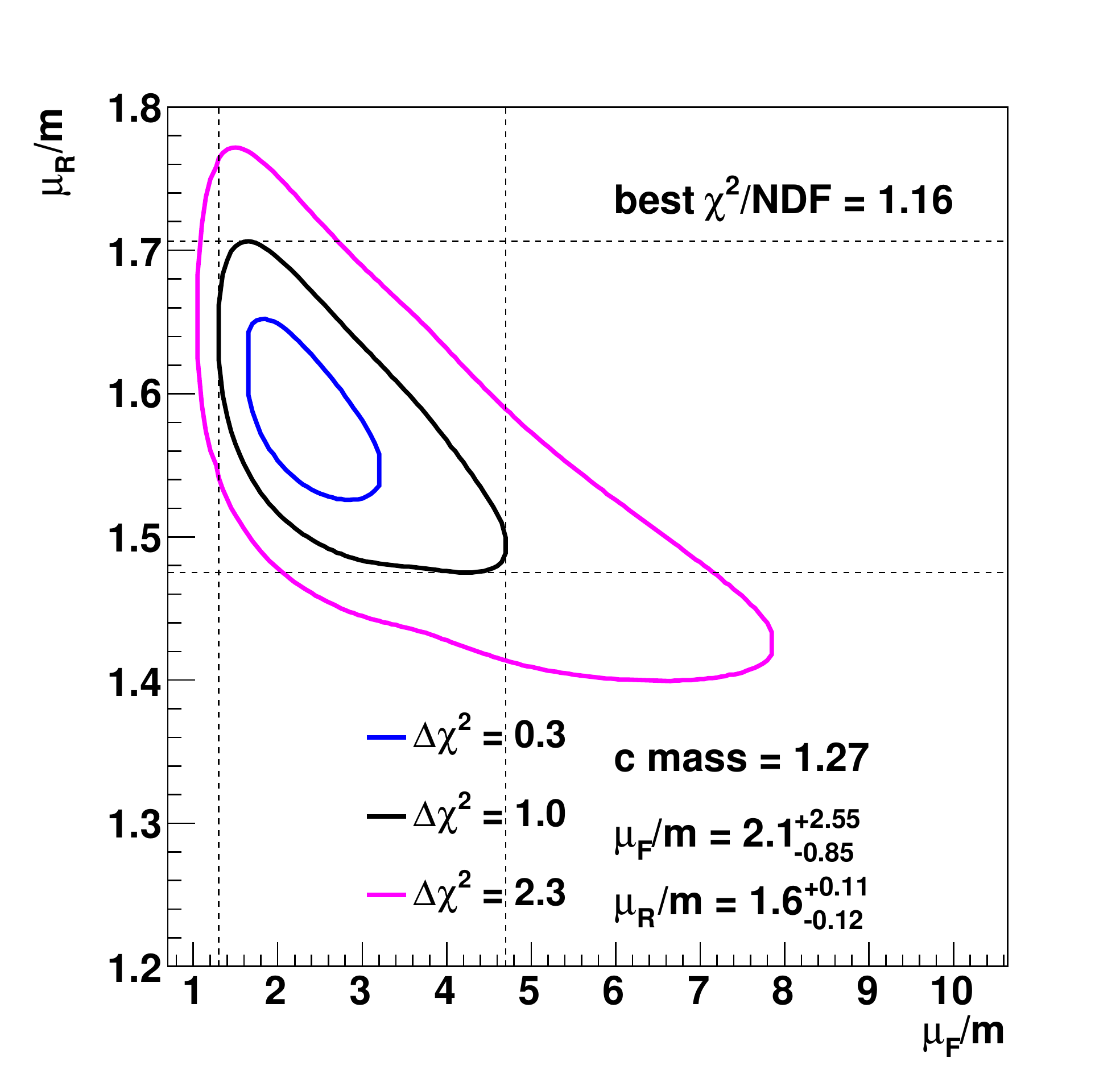} 
\includegraphics[width=0.245\textwidth,height=0.25\textheight]{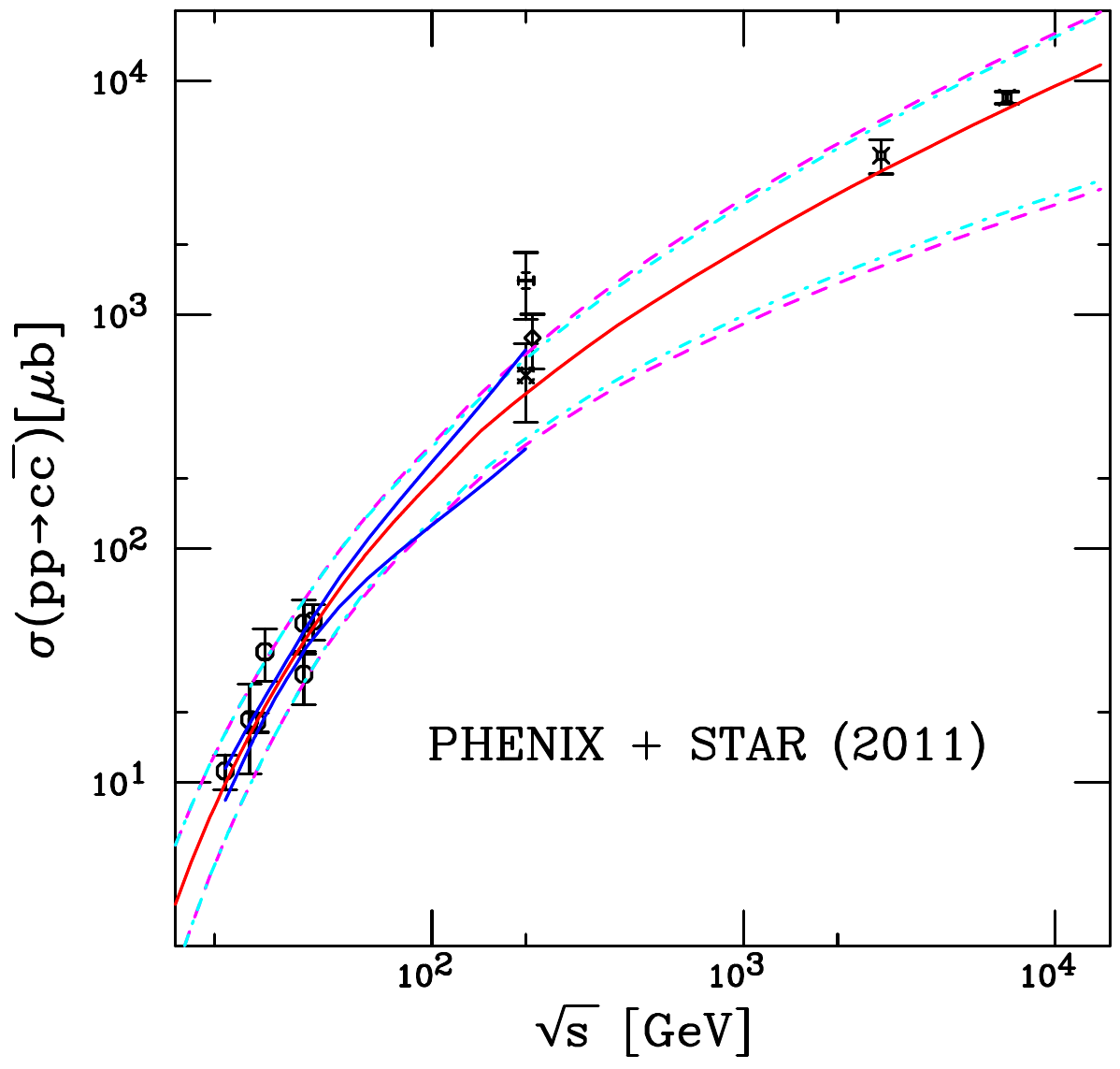} 
\includegraphics[width=0.245\textwidth,height=0.25\textheight]{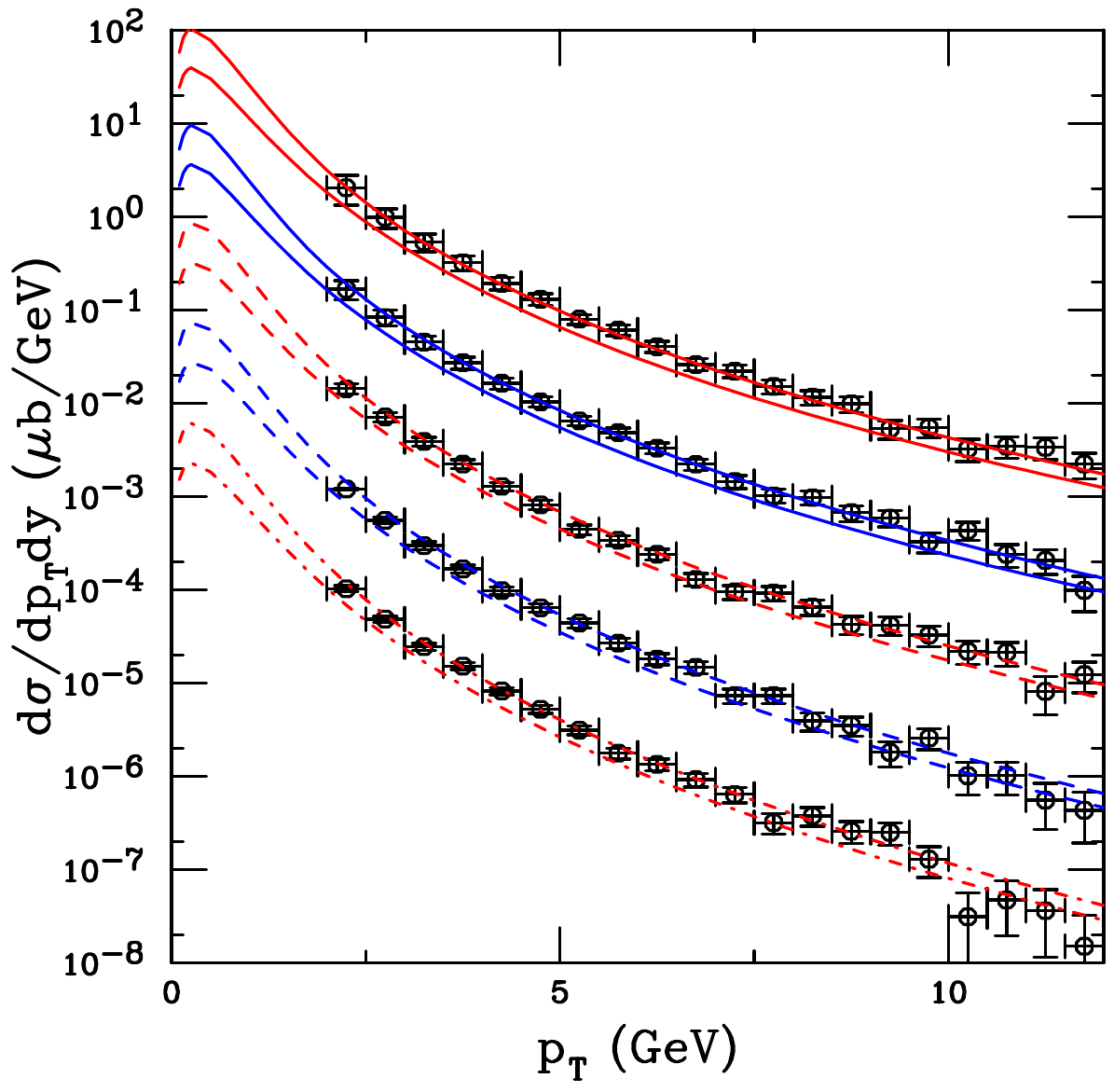} 
\includegraphics[width=0.245\textwidth,height=0.25\textheight]{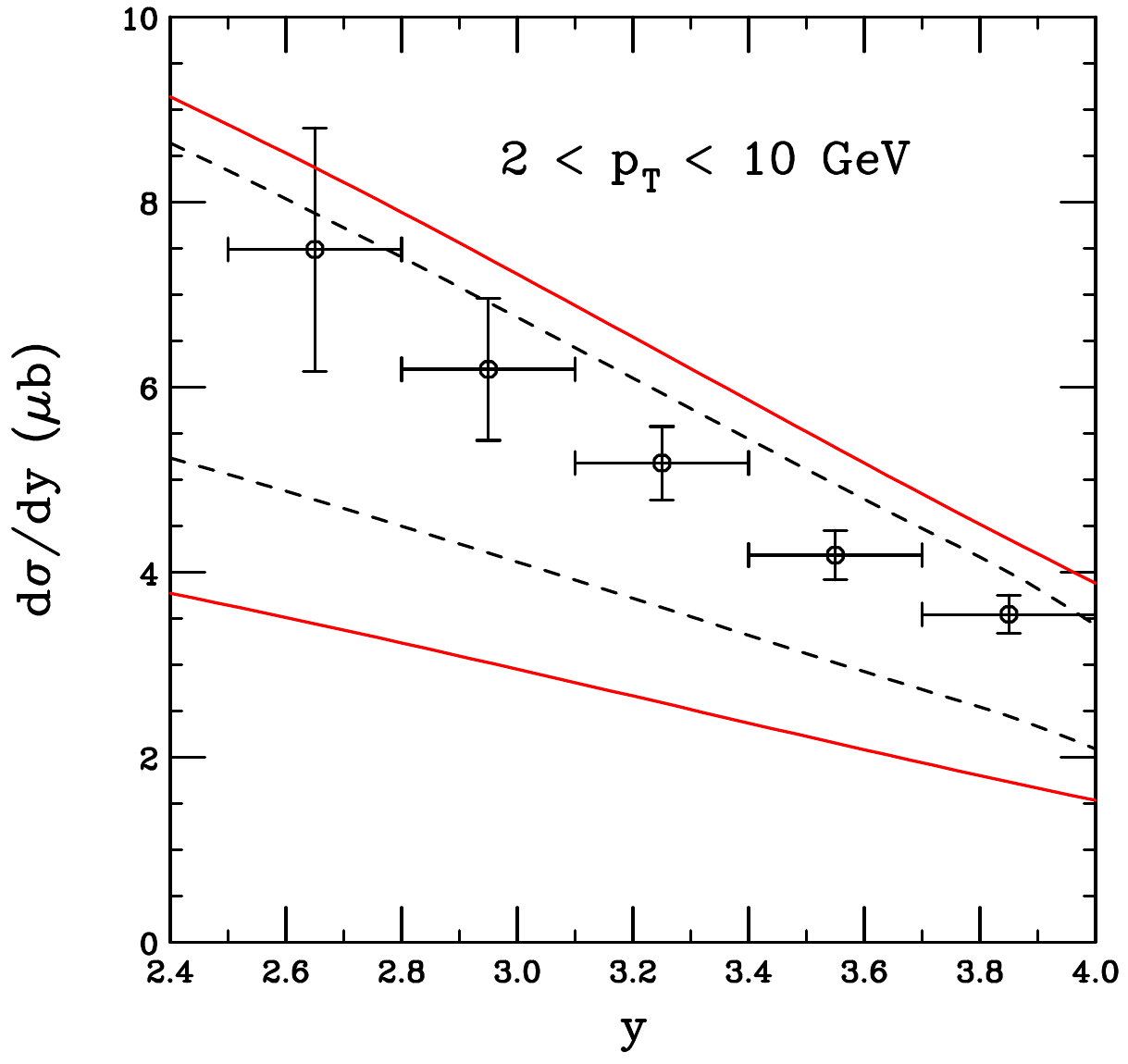} 
\end{center}
\caption[]{(Left) The $\chi^2$/dof contours for a fit to the fixed target data 
and the PHENIX and STAR 2011 cross sections at $\sqrt{s} =200$ GeV. 
The best fit values are given for
the furthest extent of the $\Delta \chi^2 = 1$ contours.
(Center left) The energy dependence of the charm total cross section compared
to data.  The central value of the fit is given by the solid red curve
while the dashed magenta curves and dot-dashed cyan curves show the extent of
the uncertainty bands, corresponding to the furthest extent of the $\Delta 
\chi^2 = 1$ contours, see text for details.  The solid blue 
curves in the range
$19.4 \leq \sqrt{s} \leq 200$ GeV represent the uncertainty obtained from the
extent of the $\Delta \chi^2 = 2.3$ contour. 
Our calculations are compared with the ALICE inclusive
single muon data from heavy flavor decays \protect\cite{ALICEmuons} at the
LHC for $pp$ collisions at
$\sqrt{s} = 7$ TeV in the center right and right panels. 
(Center right) The contributions to the $p_T$ distributions
in (a) divided into rapidity bins, from top to bottom: $2.5 < y < 2.8$ (solid
red); $2.8 < y < 3.1$ (solid blue); $3.1 < y < 3.4$ (dashed red); $3.4 < y <
3.7$ (dashed blue); and $3.7 < y < 4$ (dot-dashed red).  The top curves are
shown at their calculated value, the others are scaled down by successive 
factors of 10 to separate them. 
(Right) The sum of contributions to the rapidity distribution
are compared with the FONLL parameter set for charm centered at $m = 1.5$ GeV 
(solid red) and our results with $m = 1.27$ GeV (dashed black).  
}
\label{fig1}
\end{figure}

The best fit, including the STAR data presented at Quark Matter 2011 
\cite{STAR11}, yields the parameter values
$m_c = 1.27$ GeV, $\mu_F/m = 2.1^{+2.55}_{-0.85}$ and $\mu_R/m 
= 1.6^{+0.11}_{-0.12}$\footnotetext{Using the final STAR data point 
\cite{STAR11final} in the fitting changes the 
upper and lower limits on $\mu_F/m$ by 8\% and 4\%
respectively, while the limits on $\mu_R/m$ change by less than 1\%.}.
We show the $\chi^2$/dof fit contours on the left-hand side of 
Fig.~\ref{fig1} for $\Delta \chi^2/{\rm dof} = 
0.3$, 1 and 2.3.  The one standard deviation uncertainty in the fitted value 
of $\mu_F/m$ ($\mu_R/m$) was taken as
the maximum extent of the $\Delta \chi^2/{\rm dof} = 1$ contour along the 
$\mu_F/m$ ($\mu_R/m$) axis.  The one standard deviation 
uncertainty in the total cross section is the range of cross sections resulting
from all combinations of $\mu_F/m$ and $\mu_R/m$ contained within the 
$\Delta \chi^2/{\rm dof} = 2.3$ contour.  The $\Delta \chi^2/{\rm dof} = 0.3$ 
contour is to guide the eye.
Note the narrow range in $\mu_R/m$ relative to the much broader $\mu_F/m$ range.
The uncertainty on $\mu_F/m$ is larger and very asymmetric.  There is a greater
uncertainty on the upper limit than the lower limit because there is
a much greater change in $xg(x,\mu_F^2)$ at lower factorization scales than when
$\mu_F \gg \mu_0$, the minimum scale of the parton densities.

The center left 
panel of Fig.~\ref{fig1} shows the energy dependence of the total
charm cross section for the fits with the corresponding
uncertainty based on results using the one standard deviation uncertainties on
the quark mass and scale parameters.  If the central, upper and lower limits
of $\mu_{R,F}/m$ are denoted as $C$, $H$, and $L$ respectively, then the seven
sets corresponding to the scale uncertainty are  $\{(\mu_F/m,\mu_F/m)\}$ =
\{$(C,C)$, $(H,H)$, $(L,L)$, $(C,L)$, $(L,C)$, $(C,H)$, $(H,C)$\}.  The
upper and lower limits on the PDG value of the charm quark mass are 1.36 and 
1.18 GeV.  The uncertainty band can be obtained for the best fit sets by
adding the uncertainties from the mass and scale variations in 
quadrature. The envelope containing the resulting curves,
\begin{eqnarray}
\sigma_{\rm max} & = & \sigma_{\rm cent}
+ \sqrt{(\sigma_{\mu ,{\rm max}} - \sigma_{\rm cent})^2
+ (\sigma_{m, {\rm max}} - \sigma_{\rm cent})^2} \, \, , \label{sigmax} \\
\sigma_{\rm min} & = & \sigma_{\rm cent}
- \sqrt{(\sigma_{\mu ,{\rm min}} - \sigma_{\rm cent})^2
+ (\sigma_{m, {\rm min}} - \sigma_{\rm cent})^2} \, \, , \label{sigmin}
\end{eqnarray}
defines the uncertainty.  The uncertainty bands are
shown for two cases: the fiducial region delineated above, similar to
Ref.~\cite{CNV}, and including the most
extreme cases $(\mu_F/m,\mu_R/m) = (H,L)$ and $(L,H)$.  The difference between
the outer magenta curves, which include these extremes, and the cyan curves,
which do not, is very small.  Therefore, it is reasonable to neglect the
extremes.  We also show the
result for a one standard deviation uncertainty in the total cross section
obtained from the $\Delta \chi^2 = 2.3$ contour in the blue lines.
We have also added the 2.76 and 7 TeV total cross sections obtained by the
ALICE collaboration in $pp$ collisions \cite{ALICEpp}, not
included in our fits.  The calculations are in rather good agreement with
the data.

We use the FONLL approach \cite{CNV} to calculate the heavy flavor
semileptonic decay kinematic distributions to compare to single lepton spectra
which include $B$ decays as well as $D$ decays.  The $B \rightarrow \mu$ and
$B \rightarrow D \rightarrow \mu$ bands are calculated with the same fiducial 
set of parameters as in Ref.~\cite{CNV}.  The $D \rightarrow \mu$ band is
calculated for our best fit parameter set.
The center right and right panels of 
Fig.~\ref{fig1} compares our calculations with the ALICE single
muon data in the forward rapidity region, $2.5 < y < 4$ \cite{ALICEmuons}. 
The data are given for $2 < p_T < 12$ GeV and separated into five rapidity 
bins, each 0.3
units wide, as shown in the center right panel of Fig.~\ref{fig1}.  The 
calculations agree well with
the measurements over the entire $p_T$ range.
On the right-hand side of Fig.~\ref{fig1} we present our results in the dashed
curves as a function
of rapidity, integrated over the same $p_T$ range as the data, $2 \leq p_T \leq
10$ GeV.  We also show the rapidity distribution obtained using the FONLL
charm parameter set with a central charm quark mass of $m = 1.5$ GeV in red.
The $p_T$-integrated ALICE data agree well with both calculations.  The
results with the fitted charm parameter set narrow the uncertainty band
without sacrificing consistency with the measured data.

We now turn to a treatment of quarkonium production within this same framework.
In the CEM, the quarkonium 
production cross section is some fraction, $F_C$, of 
all $Q \overline Q$ pairs below the $H \overline H$ threshold where $H$ is
the lowest mass heavy-flavor hadron.  
We fit $F_C$ to the forward (integrated over $x_F > 0$) 
$J/\psi$ cross section data on only $p$, Be, Li,
C, and Si targets.  In this way, we avoid uncertainties due to 
ignoring any cold nuclear matter effects which are on the order of a few percent
in light targets.  We also restricted ourselves to the forward cross sections
only.

\begin{figure}[htb]
\begin{center}
\includegraphics[width=0.245\textwidth,height=0.25\textheight]{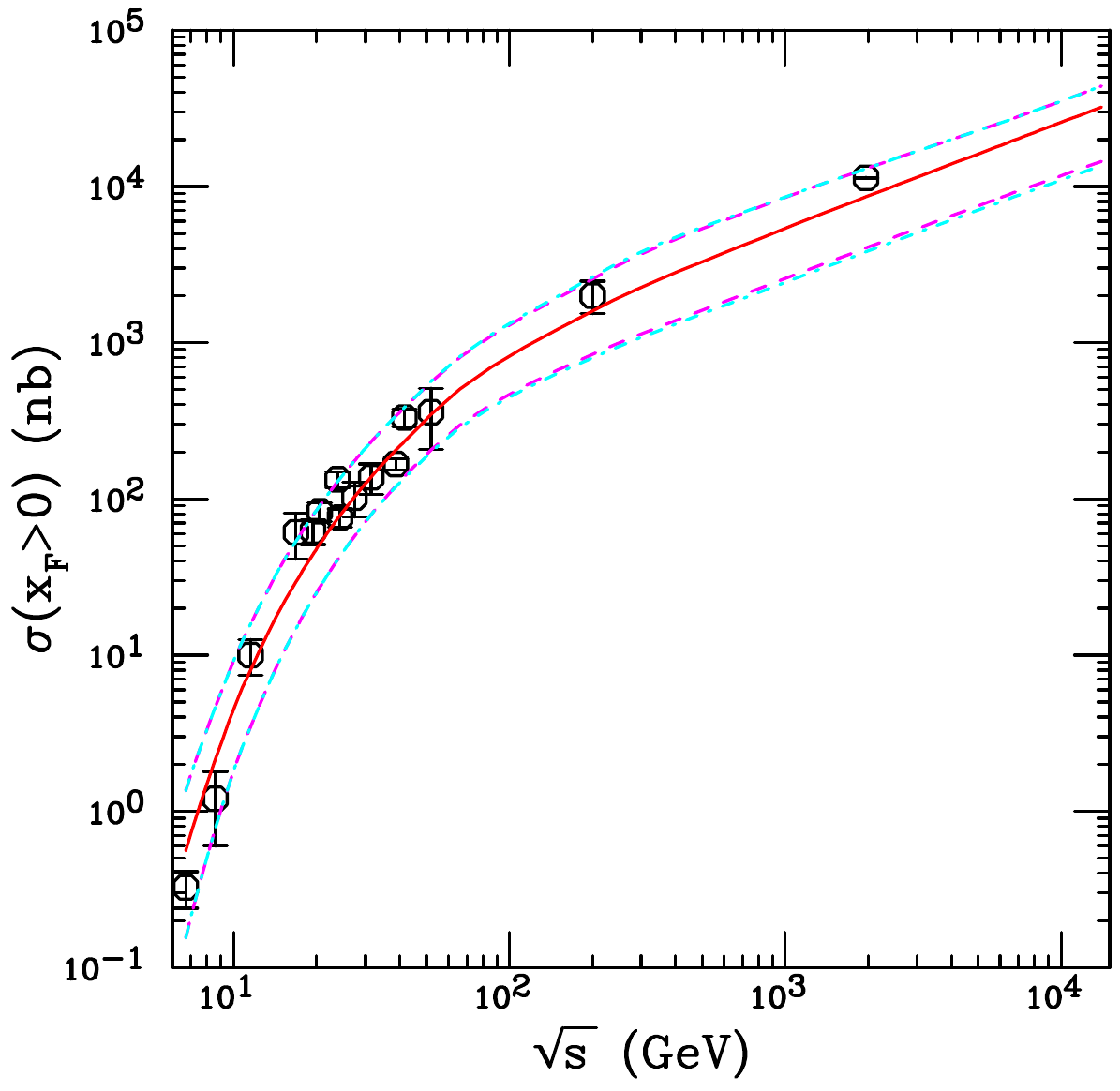} 
\includegraphics[width=0.245\textwidth,height=0.25\textheight]{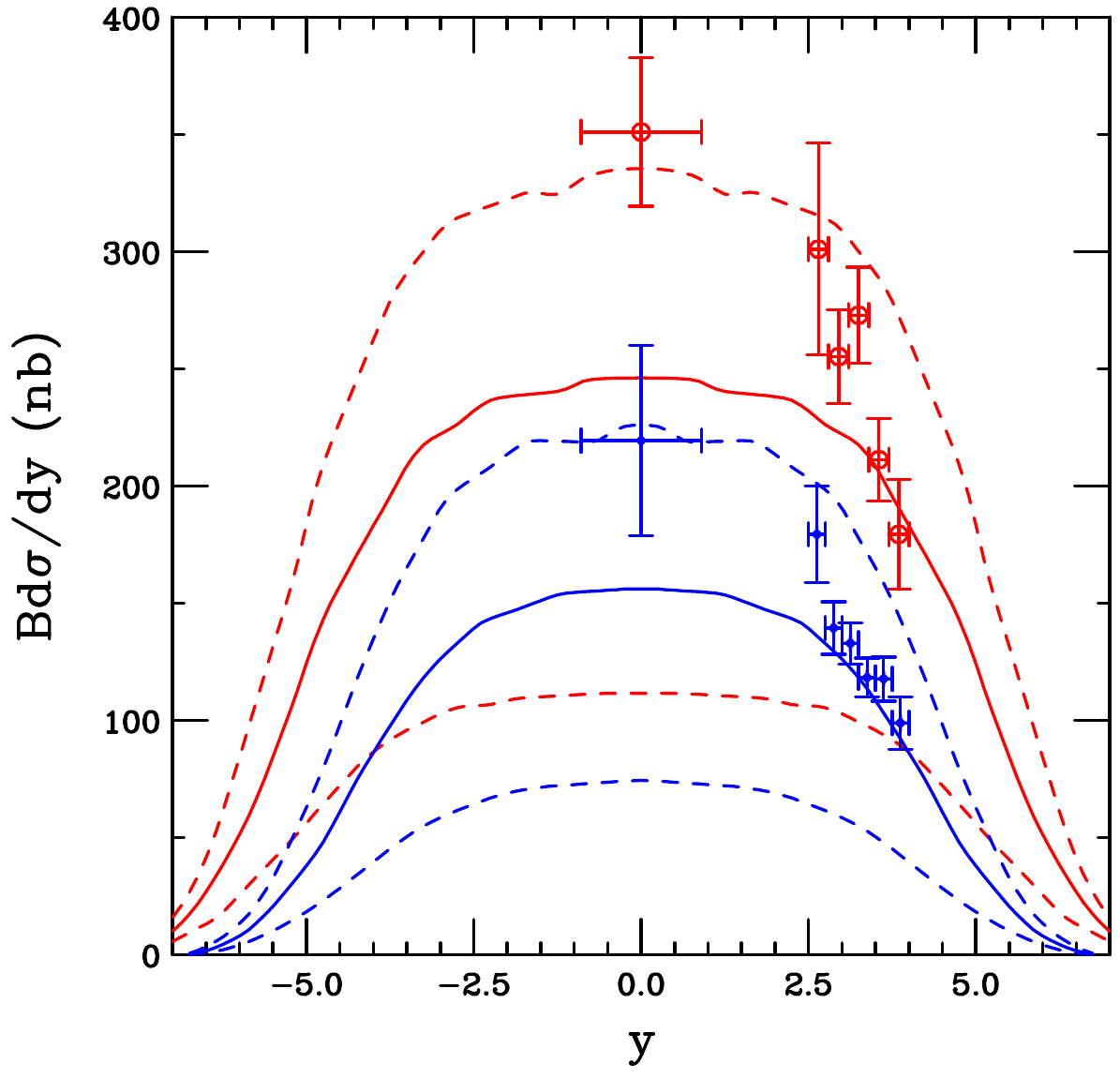} 
\includegraphics[width=0.245\textwidth,height=0.25\textheight]{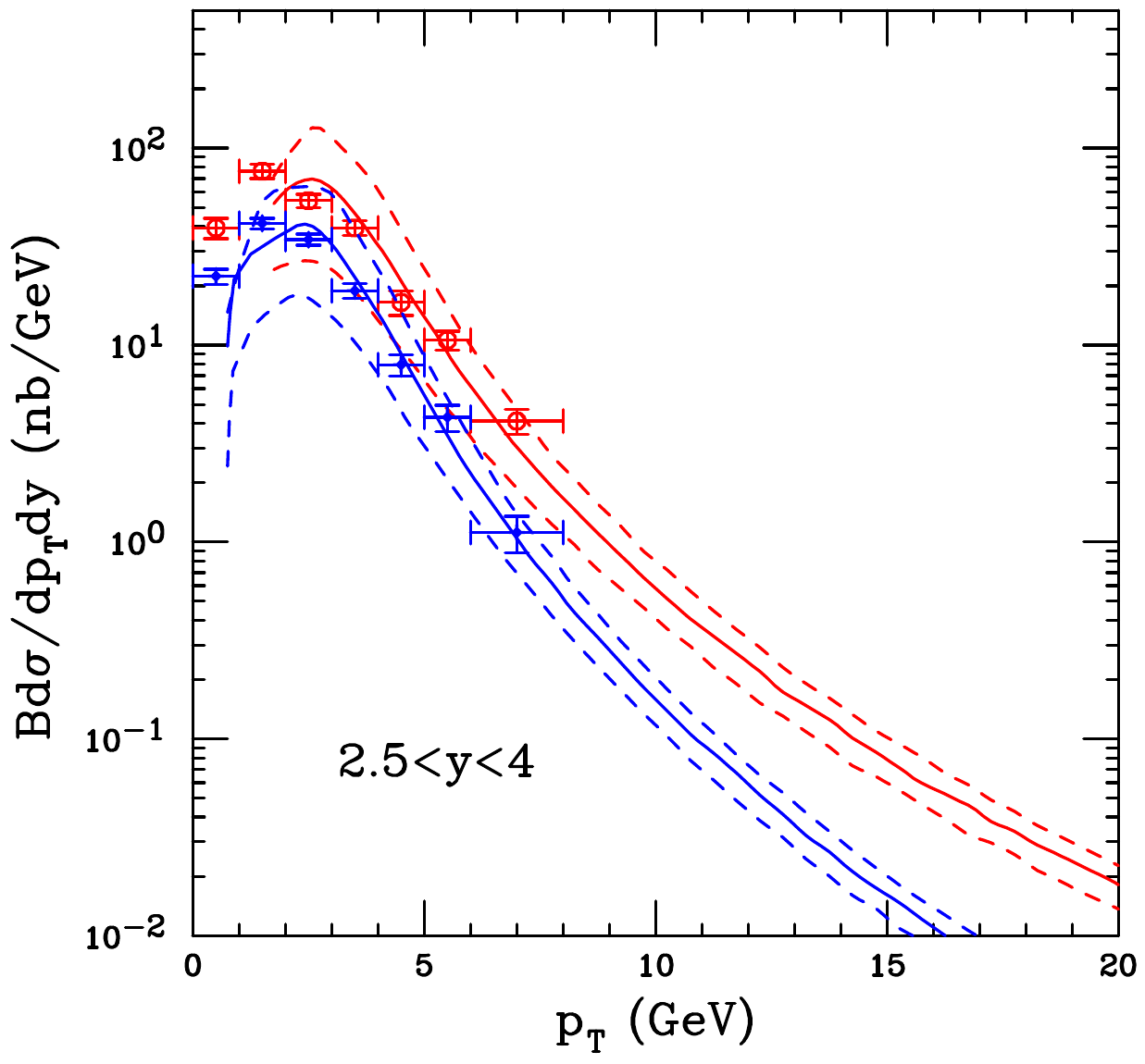} 
\includegraphics[width=0.245\textwidth,height=0.25\textheight]{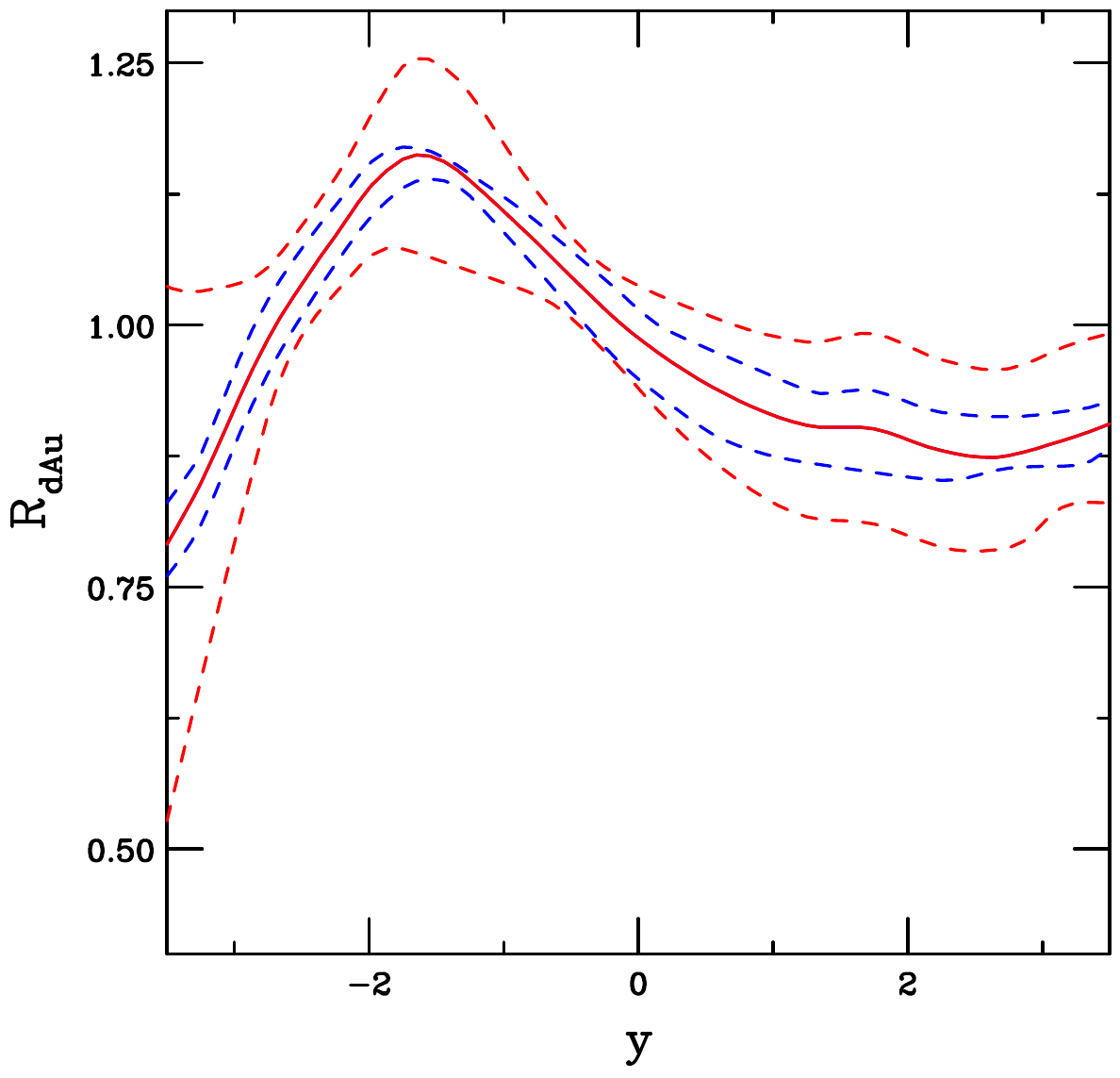} 
\end{center}
\caption{(Left) The uncertainty band on the forward $J/\psi$ cross section 
calculated based on the $c \overline c$ fit.
The dashed magenta curves and dot-dashed cyan curves show the extent of
the corresponding uncertainty bands.  The dashed curves outline the most extreme
limits of the band. 
(Center left) The $J/\psi$ rapidity distributions compared to data from 7 TeV 
\protect\cite{ALICEpp7TeV}
(red points and band) and 2.76 TeV \protect\cite{ALICEpp276TeV} 
(blue points and band).
(Center right) The forward $p_T$ distributions ($2.5 < y < 4$) are also shown.
No additional scaling factor
has been applied.  
A $\langle k_T^2 \rangle$ kick of 1.49 GeV$^2$ (7 TeV) and 1.41 GeV$^2$ (2.76
TeV) is applied to the $p_T$ 
distributions.
(Right) The scale variation of $R_{\rm dAu}$ with the central EPS09 set (blue)
compared to the EPS09 variation for the central parameter set (red).}
\label{fig2}
\end{figure}

We use the same values of the central charm quark
mass and scale parameters as we found for open charm 
to obtain the normalization $F_C$ for
$(m,\mu_F/m, \mu_R/m) = (1.27 \, {\rm GeV}, 2.1,1.6)$).  
We determine $F_C$ only for the central parameter set and scale
all the other calculations by the same value of $F_C$ to
obtain the extent of the $J/\psi$ uncertainty band.  The result is shown
on the left-hand side of Fig.~\ref{fig2}.

The ALICE 2.76 and 7 TeV inclusive $J/\psi$ rapidity and forward 
$p_T$ distributions ($2.5 \leq y \leq 4$) are shown in 
the center left and center right plots of Fig.~\ref{fig2}.
The rapidity distribution at $\sqrt{s} = 7$ TeV
is flat over several units of rapidity.  The calculated rapidity distribution 
at 2.76 TeV is not as broad and the agreement with the data is rather good 
although the 
midrapidity point remains high relative to the central value of the calculation.
The agreement of the calculated $p_T$ distributions with the forward rapidity
data is quite good with the exception of the lowest $p_T$ points where the
calculated distributions turn over more quickly than the data.

Finally, in the right-hand panel of Fig.~\ref{fig2}, we show the ratio
$R_{\rm dAu}$ for 200 GeV d+Au collisions at RHIC to NLO in the total cross
section with EPS09 NLO shadowing.  The red band shows the variation with
respect to the 31 EPS09 sets while the blue band indicates the mass and scale
variation around the central set.  We note that, with the best fit set, the
scale uncertainty is now less than the uncertainty due to shadowing.

We have narrowed the uncertainty band on the open heavy flavor cross section
and, in so doing, have also provided a realistic uncertainty band on $J/\psi$
production in the Color Evaporation Model.  The central result, $m = 1.27$ GeV,
$\mu_F/m = 2.1$ and $\mu_R/m = 1.6$, is quite compatible with previous
calculations using a `by-eye' fit to the data with $m = 1.2$ GeV, $\mu_F/m =
\mu_R/m = 2$ \cite{Gavai:1994in,vogtHPC}. 

While the fits have been made by comparing the calculated NLO charm production
cross section to available data at fixed-target energies and at RHIC, they are
in good agreement with the extracted total charm cross sections at the LHC.
The same parameter set also provides good agreement with the distributions of
single leptons from semileptonic heavy flavor decays at RHIC and the LHC.
The limit on the width of the uncertainty band is now set by the uncertainty
due to bottom quark production and decay.

We have used the same fit parameters in the calculation of $J/\psi$ production
in the color evaporation model and have thus provided the first uncertainty
band on $J/\psi$ production in this approach.  The energy dependence of the
total $J/\psi$ cross section that results is a good match to the data up to
collider energies.  The $p_T$ distributions are also in good agreement with
the data from RHIC and the LHC. In future work, we will use our new parameter
set to place limits on the contribution of $B$ meson decays to $J/\psi$ 
production and will also study cold nuclear matter effects on $J/\psi$ 
production in more detail at next-to-leading order.

\section*{Acknowledgments}

We thank M. Cheng, L. Linden Levy, P. Petreczky, R. Soltz 
and P. Vranas for discussions.
The work of R. V. and R. E. N. was performed under the auspices of the U.S.\
Department of Energy by Lawrence Livermore National Laboratory under
Contract DE-AC52-07NA27344 and was also supported in part by the
JET Collaboration. The work of A. D. F. was supported by the National
Science Foundation grant PHY-07-54674.








\end{document}